# Modelling the non-linear dynamics of the looping pendulum

# Research Report

Avighna Daruka*

*The Doon School, Dehradun India*

*Mr. Gyaneshwaran Gomathinayagam*

*Mr. Aneesh Agarwal*

## Abstract

The Looping pendulum phenomenon was first introduced in 2019 at the 32nd edition of the IYPT, wherein a lighter bob sweeps around a cylindrical rod to support the weight of a heavier bob. In this paper, the phenomenon was divided based on rotating and non-rotating forces, and differential equations were derived for each. To verify the theoretical derivation, an experimental analysis was done, varying the mass ratio with the vertical distance travelled by the heavier bob. (Tracked using tracker) Experimental findings fit a logarithmic curve fit – falling succinctly with a similar trend with the simulation run with MATLAB solving the derived differential equations. Furthermore, to verify the simulation, the trajectory of both the lighter and heavier mass was also compared for the simulation and experimental findings. The experimental findings fit very closely to the simulation findings, accrediting the validity and accuracy of the derived theory.

# Table of Contents

# 1.0 Introduction

In the Looping Pendulum Phenomenon, a small bob *m* sweeps around a cylindrical rod in a spiral manner to hold the weight of the heavier bob *M*. In this paper, I aim to explore this interplay of applied Newtonian Mechanics and then work on the derivation of novel equations of motion representing this phenomenon, henceforth solving this problem proposed by the IYPT in 2019 (1). *(Figure 1)*

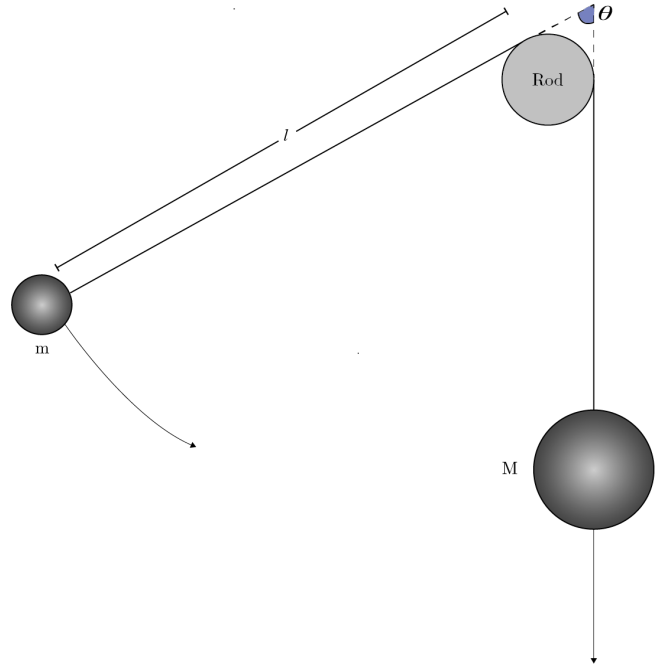

Figure 1: Diagrammatic Representation of the Looping Pendulum; $m$ makes an angle of $\theta$ with the vertical. The length of the string by which $m$ is suspended from the rod is $l$. The two arrows signify the direction of motion of the two bobs before $M$ is held up.

This theoretical model derivation can give us deeper insights on nonlinear harmonic oscillation and can help us understand the underlying's of chaotic behavior and numerical solutions on real life phenomena. This can then be used to develop algorithms for operating robotic arms and other articulated devices that behave similarly (*not discussed in this paper*). This is only one of many examples of how knowing the workings of the looping pendulum may be applied in real life to extract solutions to chaotic circumstances where little changes in parameters have a significant influence on the final variable being investigated.

C. Dannheim et. al. (4) proposed in their theoretical derivation of the looping pendulum that the friction between the string and the rod increases exponentially with the angle of wrapping of the string. This is intuitive in nature as more of the string wrapping around the rod (and hence the angle increasing) would mean more of the surface between the string and rod coming in contact - leading to the friction increasing in a similar manner. However, their derivation is only valid for a massless string, whereas the derivation here generalizes this to a string with non-negligible mass (and thus, can describe a looping pendulum without $m$, as its mass doesn't need to be accounted for).

## 2.0 Theoretical Model Derivation

### 2.1 Intuitive Explanation of Acting Forces

The theoretical model was derived by dividing the phenomenon into two parts. *(Figure 2)* Rotating (lighter bob sweeping around the rod), and non-rotating (heavier bob falling vertically). In the latter, two forces are acting, the weight of the heavier pendulum bob (named $M$) and the tension of the string following it. As mentioned before, previously published research (4) predicts that the friction of the string following the heavier pendulum bob increases exponentially as the lighter bob wraps around the rod. As the contact surface area also increases in a similar manner, the tension in the string following $M$ must similarly increase due to the string being pulled in two directions. This has been proved algebraically using the Capstan Equation. (5)

In the rotating component of the Looping Pendulum, there are also two forces. The gravitational force in the downward direction, and the tension along the length of the string. The tension along the length of the string does not exert any torque on the lighter bob (denoted as $m$) as the line of action of tension force which is acting on the lighter bob passes through its own axis of rotation and hence, cannot exert any torque as the moment of inertia would be zero. *(Figure 3)*

We analyze all these forces in greater detail in the upcoming sections of the paper.

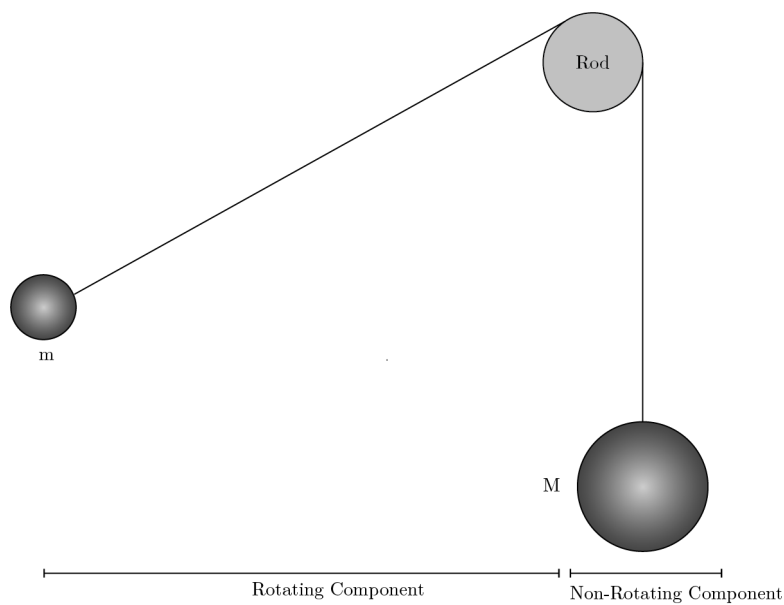

Figure 2: Classification of the Looping Pendulum Phenomenon into two components, the rotating and non-rotating components.

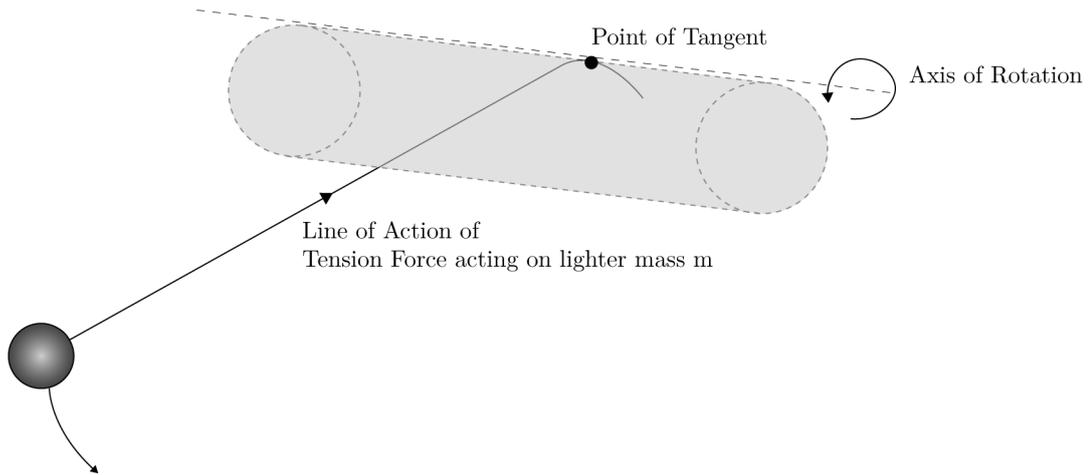

Figure 3: The Line of Action of Tension force intersects the Axis of Rotation, and hence does not exert any torque on the smaller bob $m$ which is rotating.

## 2.2 Rotating Component of Looping Pendulum

### 2.2.1 Sign Convention and Definitions

#### 2.2.1.1 Coordinate System

As shown on Figure 3, the axis of rotation is the point where the string following bob $m$ is tangent to the rod. As the lighter bob sweeps around the rod, it would be tangent at different directions and hence, there will be a moving origin as shown in Figure 4.

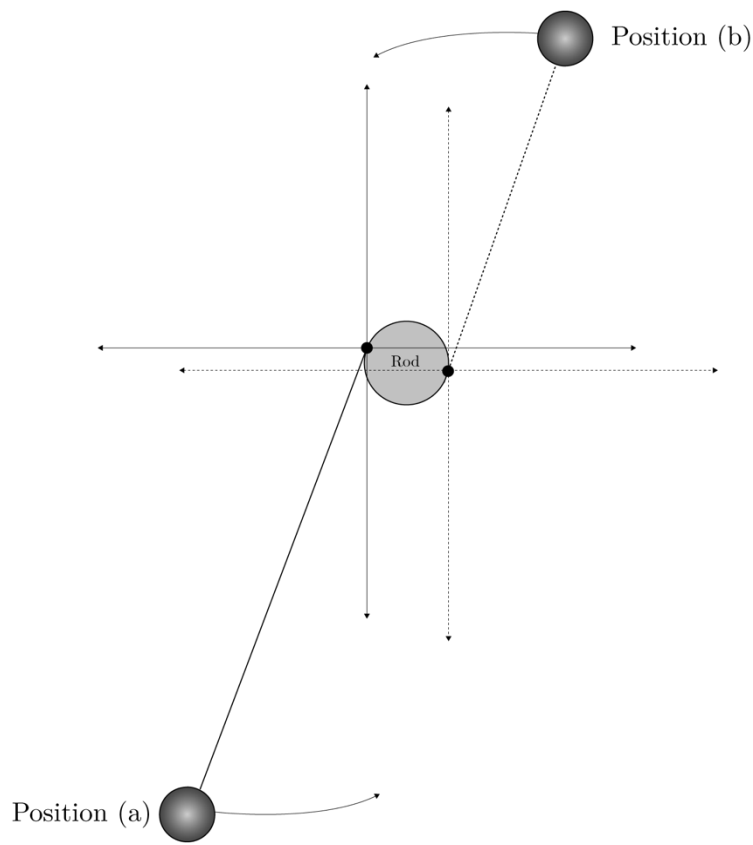

Figure 4: The axis of rotation is the point at which the string is tangent to the metal rod, this is variable and changing as the lighter bob sweeps around the rod.

### 2.2.1.2 Angle Nomenclature

This paper adopts the anticlockwise direction as positive, and the angle θ that *m* makes with the vertical is defined to be taken from the -y axis. This is visualized in Figure 5.

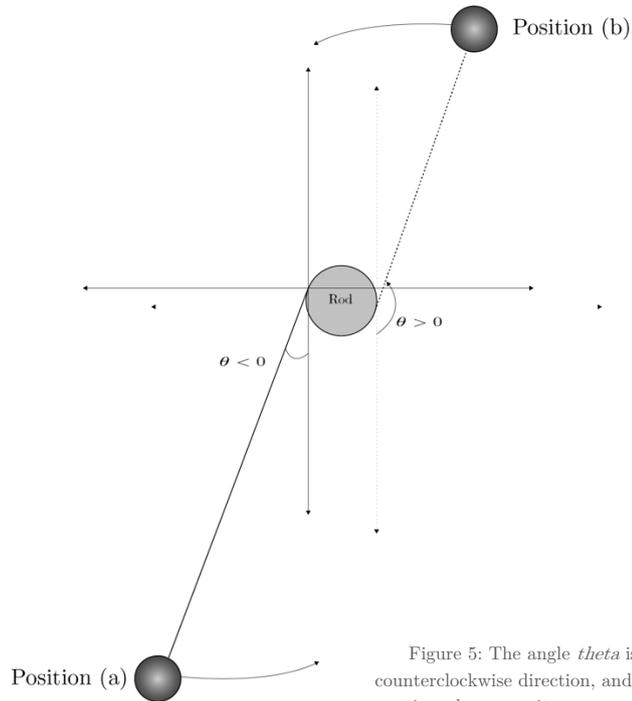

Figure 5: The angle *theta* is defined in the counterclockwise direction, and hence initially it is negative whereas as it sweeps around the angle with the -y axis it is positive.

As *m* rotates around the rod, the angle increases from the initial value $\theta_0$ (which is negative) to 0 (when it first becomes vertical) and remains positive thereafter (Figure 6).

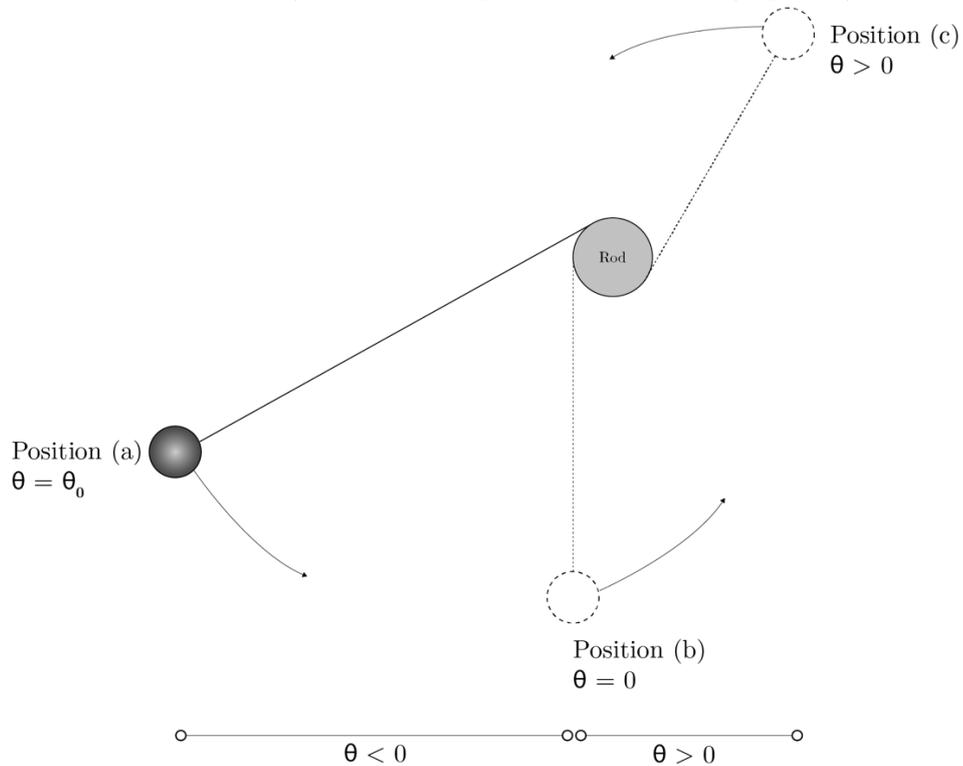

Figure 6: The Angle progression as M sweeps around the rod using the newly established nomenclature.

### 2.2.2 Angular Motion of Rotating Part

#### 2.2.2.1 Torque Derivation

The angle between the radius and weight vectors has been defined as $\theta$ as shown in figure 7. The torque exerted by the string (mass $M_s$) is assumed to be distributed equally throughout the mass of the string, but is averaged at the center of the string, and hence its torque derivation has been divided by two. Thus, the magnitude of the total torque acting upon the rotating part of the pendulum is:

$$|\tau| = \frac{M_s gl \cdot sin(\theta)}{2} + mgl \cdot sin(\theta)$$

...........(1)

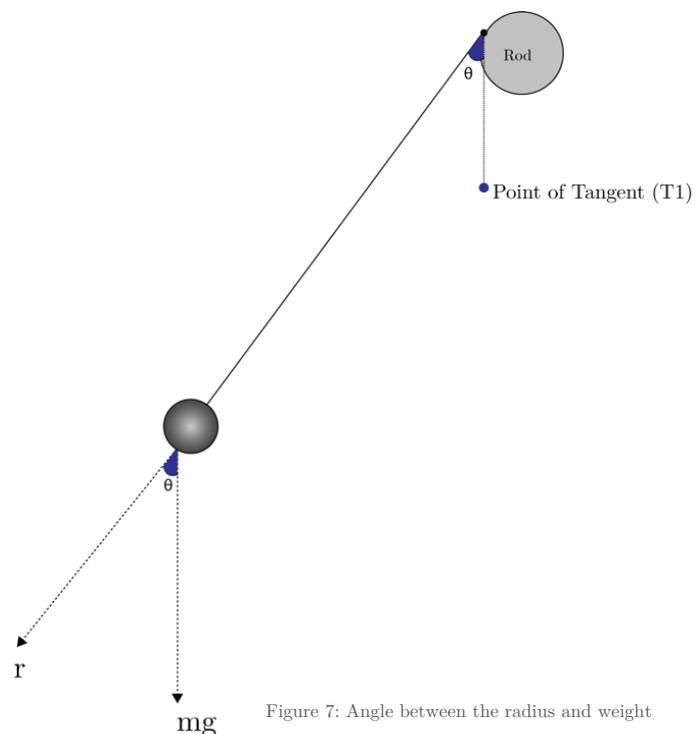

Figure 7: Angle between the radius and weight vectors has been defined as theta.

The modulus sign does not account for the anticlockwise and clockwise torques. And hence the equation (1) must be multiplied with -1 to account for this.

$$\tau = -\frac{M_s gl \cdot sin(\theta)}{2} - mgl \cdot sin(\theta) \quad ...(2)$$

Since $\tau = \frac{dL}{dt}$ where (L is the angular momentum and is equal to $I\omega$)[1]

$$\tau = \frac{d(I\omega)}{dt}$$

Which, when simplified using the Differentiation product rule:

$$\tau = \frac{dI}{dt}\omega + I\frac{d\omega}{dt}$$

...(3)

---

[1] Initially, I thought that I could use the relation $\tau = I\alpha$ to arrive at a differential equation describing the motion of the rotating part. However, later, when I created the simulation to predict the trajectory of $m$ and $M$, I noticed that a spiral was never being formed. This was in direct conflict with the results of my experiment. Upon reinvestigating my theory, I realized that this expression for torque can only be used in the case of a rigid body whose moment of inertia is constant (In that case, the first term on the right-hand side in equation (3) becomes zero, resulting in $\tau = I\alpha$). However, in my case, since the length of the string was constantly reducing, the moment of inertia was also continuously decreasing. Thus, I used this form of torque, which is given by the instantaneous rate of change of angular momentum.

### 2.2.2.2 Moment of Inertia Derivation (about axis of rotation)

When the axis of rotation of a rod goes through one end of the rod, the total moment of inertia $I$ is obtained by summing the moment of inertia for the point mass and the string. (18) Given by:

$$I = \frac{M_s l^2}{3} + ml^2$$

...(4)

### 2.2.2.3 Deriving Differential Equation in respect to theta

Substituting (4) into (3)

$$\tau = \frac{d\left(\frac{M_s l^2}{3} + ml^2\right)}{dt} \omega + \left(\frac{M_s l^2}{3} + ml^2\right)\frac{d\omega}{dt}$$

...(5)

Simplification using implicit differentiation and the chain rule yields:

$$\tau = \left(\frac{2M_s l}{3} \cdot \frac{dl}{dt} + 2 \cdot ml \cdot \frac{dl}{dt}\right)\omega + \left(\frac{M_s l^2}{3} + ml^2\right)\frac{d\omega}{dt}$$

Further Simplification:

$$\Rightarrow \tau = \frac{2l\omega}{3}(M_s + 3m) \cdot \frac{dl}{dt} + \frac{l^2}{3}(M_s + 3m) \cdot \frac{d\omega}{dt}$$

When we equate equation (2) and (5):

$$-\frac{M_s g l \cdot sin(\theta)}{2} - mgl \cdot sin(\theta) = \frac{2l\omega}{3}(M_s + 3m) \cdot \frac{dl}{dt} + \frac{l^2}{3}(M_s + 3m) \cdot \frac{d\omega}{dt}$$

Simplifying Gives:

$$-\frac{gl \cdot sin(\theta)}{2}(M_s + 2m) = \frac{2l\omega}{3}(M_s + 3m) \cdot \frac{dl}{dt} + \frac{l^2}{3}(M_s + 3m) \cdot \frac{d\omega}{dt}$$

Dividing both sides by $\frac{l(M_s+3m)}{3}$ yields:

$$-\frac{3}{2}g \cdot sin(\theta)\frac{M_s + 2m}{M_s + 3m} = 2\omega\frac{dl}{dt} + l \cdot \frac{d\omega}{dt}$$

Since $\omega = \frac{d\theta}{dt}$ :

$$-\frac{3}{2} g \cdot sin(\theta) \cdot \frac{M_s + 2m}{M_s + 3m} = 2 \cdot \frac{d\theta}{dt}\frac{dl}{dt} + l \cdot \frac{d\left(\frac{d\theta}{dt}\right)}{dt}$$

$$\Rightarrow -\frac{3}{2} g \cdot sin(\theta) \cdot \frac{M_s + 2m}{M_s + 3m} = 2 \cdot \frac{d\theta}{dt}\frac{dl}{dt} + l \cdot \frac{d^2\theta}{dt^2}$$

$$\Rightarrow \frac{d^2\theta}{dt^2} = -\frac{3}{2l} g \cdot sin(\theta) \cdot \frac{M_s + 2m}{M_s + 3m} - \frac{2}{l} \cdot \frac{d\theta}{dt}\frac{dl}{dt}$$

...(6)

### 2.2.3 Radial Motion of Rotating Part

As the heavier bob moves vertically down, we can observe a radially inward motion of the lighter bob as shown in figure 8. In this radial direction, there are two forces that are acting on this lighter bob: The tension $T_r$ of the string following the heavier bob, and mg · cos ($\theta$) (Figure 8)

As both these forces are in opposing directions, we assume the radially inward vector ($T_r$) as negative. And hence the resultant force acting on the bob $m$ is:

mg · cos ($\theta$) - $T_r$

By equating,

m$a_r$ = mg · cos ($\theta$) - $T_r$

...(7)

Where $a_r$ is the radial acceleration of bob $m$.

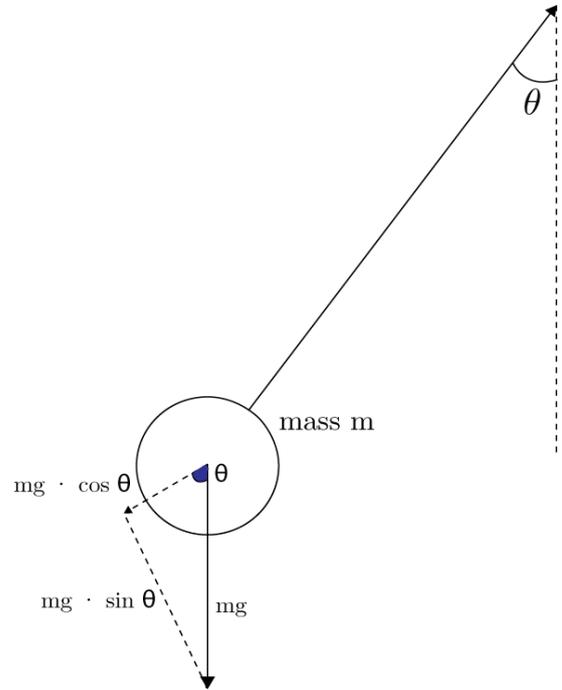

Figure 8: There are two forces acting on bob $m$ which represent the net force acting on the bob. The mg component has been resolved into its two components.

## 2.3 Non Rotating Component of Looping Pendulum

### 2.3.1 Net Force Analysis

As the heavier bob does not rotate, we can define its motion as rectilinear along the -y axis (defined before). The heavier bob starts at y=0 and proceeds downwards. (Figure 9)

A construction of a free-body diagram was then conducted on the heavier bob. (Figure 10)

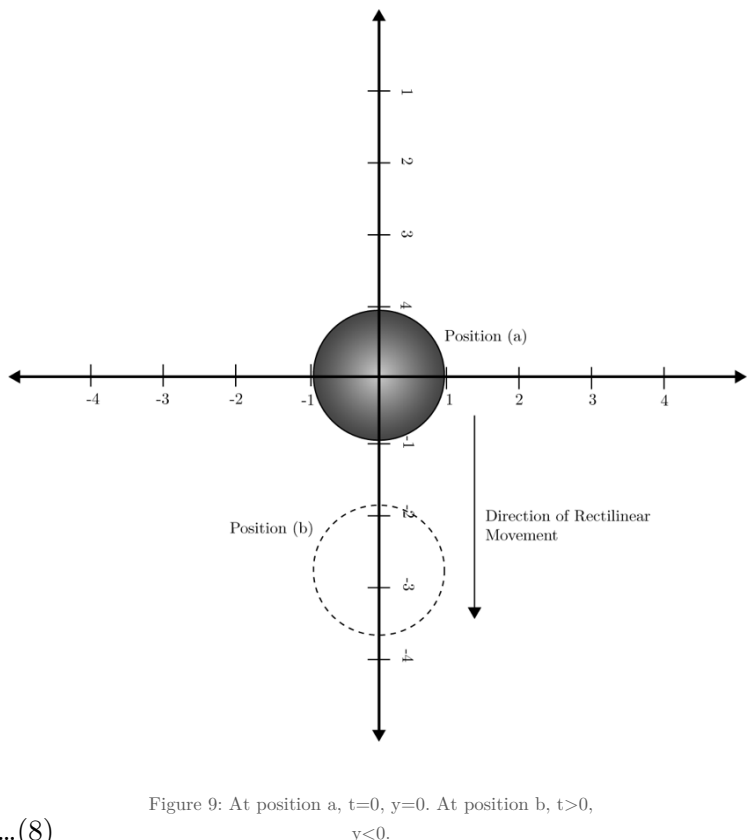

Figure 9: At position a, t=0, y=0. At position b, t>0, y<0.

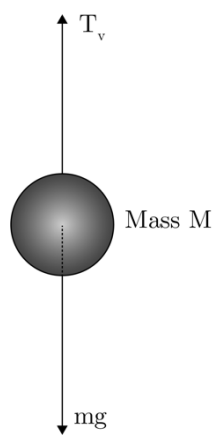

$$Ma = T_v - mg$$
Where $(a = \frac{d^2y}{dt^2})$ thus
$$\frac{d^2y}{dt^2} = \frac{T_v}{M} - g$$
...(8)

Figure 10: On the heavier bob there are two forces, the tension which acts vertically upwards, and the weight which acts vertically downwards.

### 2.3.2 Derivation of Tension $T_r$

#### 2.3.2.1 Approach to Derivation

As the rod has a circular cross section, the forces that act on infinitesimal small parts of the string act along constantly changing directions, and hence if we can construct an expression for one infinitesimally small part of the string (defined as *dm*), we can integrate it over the entire section of the string to obtain the total friction of the structure.

The string (including *dm*) has the same acceleration of $\frac{d^2y}{dt^2}$. And hence the net resultant force on *dm* is dm multiplied with $\frac{d^2y}{dt^2}$. (Figure 11).

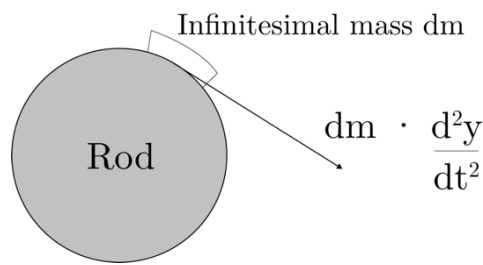

There is no net force normal to the tangential force, and hence it does not affect the value of the net force calculated.

Figure 11: The resultant force acts in the tangential direction with a value of dm multiplied with the acceleration.

### 2.3.2.2 Force analysis of *dm* on string

A free body diagram was then constructed for *dm* which has been shown in Figure 12. Along the normal to the rods surface, the normal force *dN* acts away from the centre of the rod. (The weight of *dm* opposing this in the other direction has been ignored as it can be approximated to zero). The force of friction *dF* acts along the tangent opposite to the direction of motion. Moreover, there are two additional forces of tension acting on *dm*. They are denoted as T and T + dt. The difference in magnitude *dT* acts as a small increment in tension that occurs across all the infinitesimal parts of the string wrapped around the rod. All these increments then add up to produce the difference in magnitude that exists between the two tensional forces on either side of the rod ($T_v$ and $T_r$). This is a reiteration of the Capstan equation and is the fundamental principle of how the weight of the heavier bob is held up.

As shown, the string is tangential to the endpoints of *dm* and hence we define this angle between the tangent at the center of the infinitesimal mass and the string, as d$\eta$. (Figure 12)

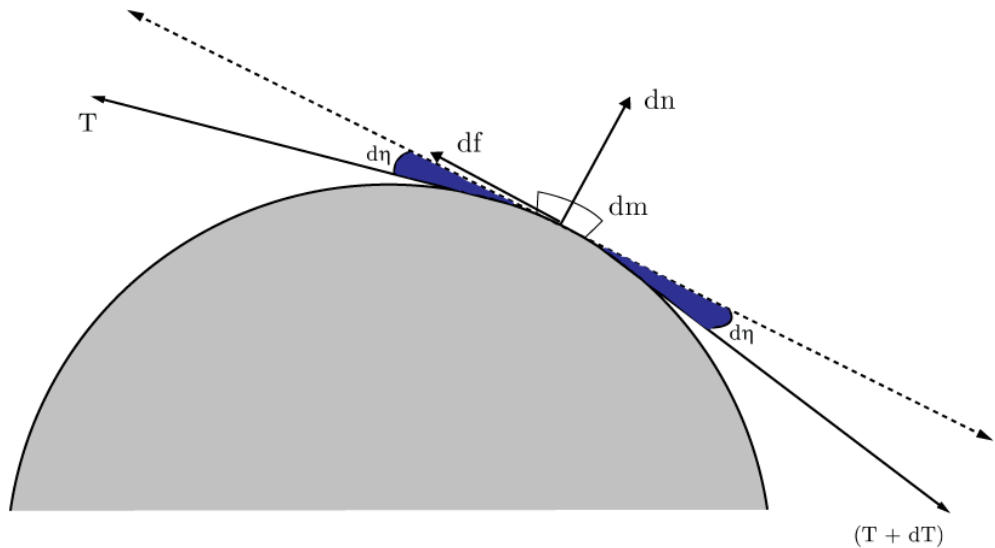

Figure 12: All the four forces acting upon the infinitesimal mass *dm* of the string have been represented in the free body diagram.

To obtain equations relating individual forces to the resultant forces in the tangential and normal directions, the tangential force was resolved into its orthogonal components along these directions. (Figure 13) (9)

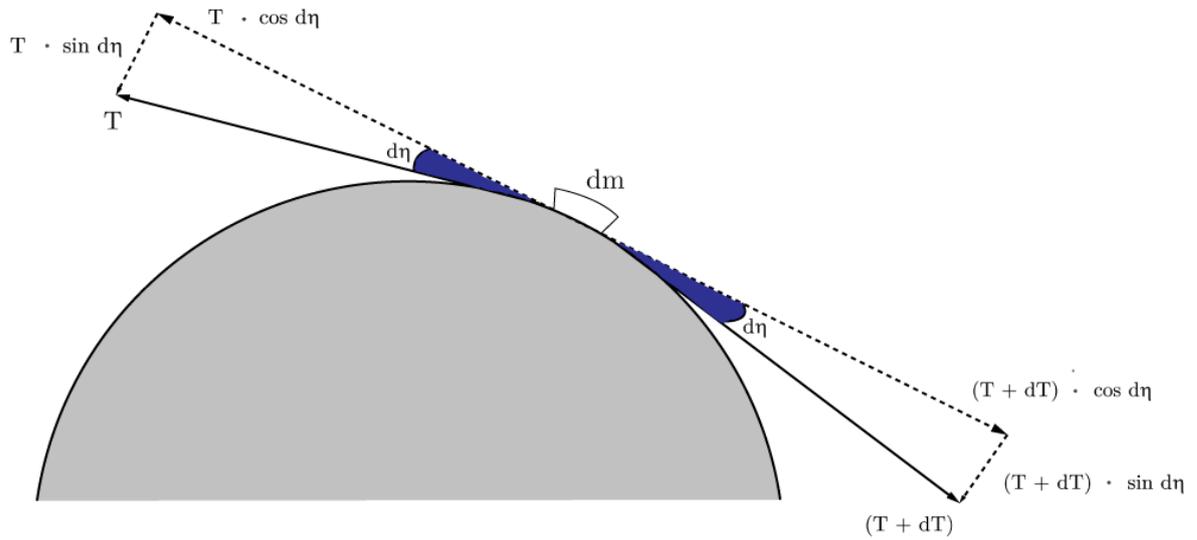

Figure 13: The forces along the tangential direction have been resolved into their orthogonal components to help derive equations relating them.

As (T · sin dη) and (T +dT · sin dη) are in the opposite direction to dN, the net force along dN is:

$$dN - (2T+dT) \cdot \sin d\eta = 0$$

According to the Taylor Series, for small angles sin (dη) ≈ dη. Therefore,

$$2T \cdot d\eta + dT \cdot d\eta = dN$$

Since we can approximate dT · dη ≈ 0:

$$2T \cdot d\eta = dN \qquad \text{...(9)}$$

Along the tangent the net force is given by dm · $\frac{d^2y}{dt^2}$ as defined in 2.3.2.1. Furthermore, as defined in 2.3.1, M is moving rectilinear along the -y axis and hence its moving in the *negative* direction, and hence, to ensure coherence, the forces leading towards M have also be defined as negative. Thus, the net force equation along the tangent is:

$$dm \cdot \frac{d^2y}{dt^2} = -dT \cdot cos(d\eta) + dF$$

According to the Taylor series, for small angles, cos(dη) ≈ 1. Hence:

$$dm \cdot \frac{d^2y}{dt^2} = -dT + dF$$

Since (dF = $\mu_d$ dN) where $\mu_d$ is the coefficient of dynamic friction between the string and the rod.

$$dm \cdot \frac{d^2y}{dt^2} = -dT + \mu_d dN \qquad \text{...(10)}$$

Substituting (9) in (10),

$$dm \cdot \frac{d^2y}{dt^2} = -dT + 2\mu_d T \cdot d\eta \qquad \text{...(11)}$$

### 2.3.2.3 Simplifying Differential Equations (11)

To integrate all the infinitesimal mass $dm$ in the portion of the wrapped string, d$\eta$ must be expressed in terms of the angle $d\o$ that $dm$ subtends at the center. (Figure 14)

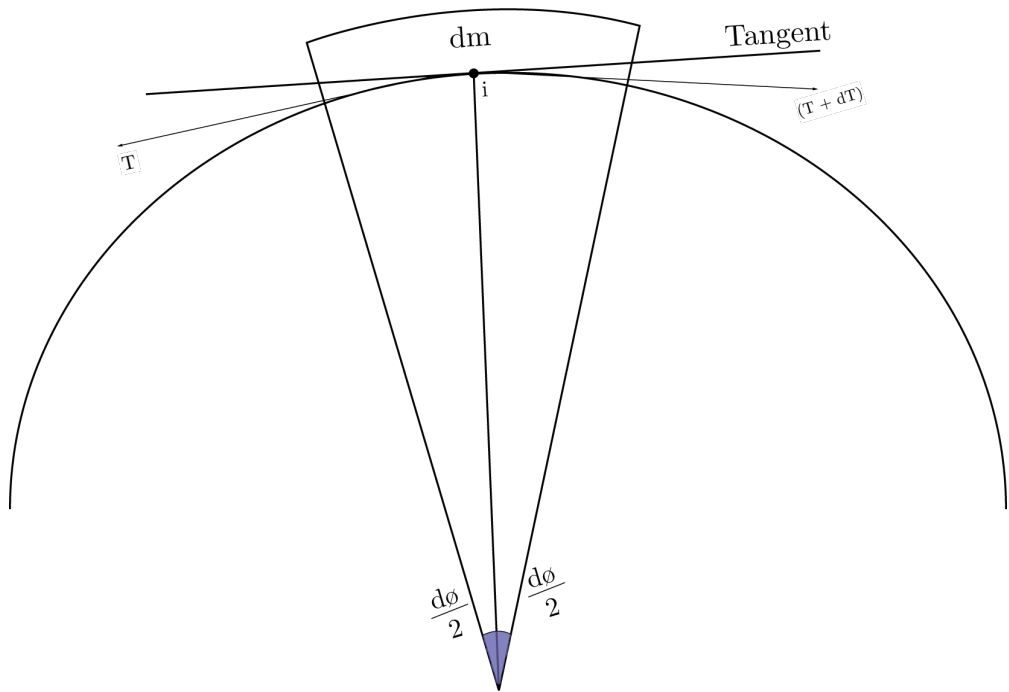

Figure 14: Since I lies at the center of dm, a line drawn from the center bisects dø.

In figure 14, we can define dη by translating (moving all the point on the line at the same amount in a given direction) the tangent to the line defined NO.

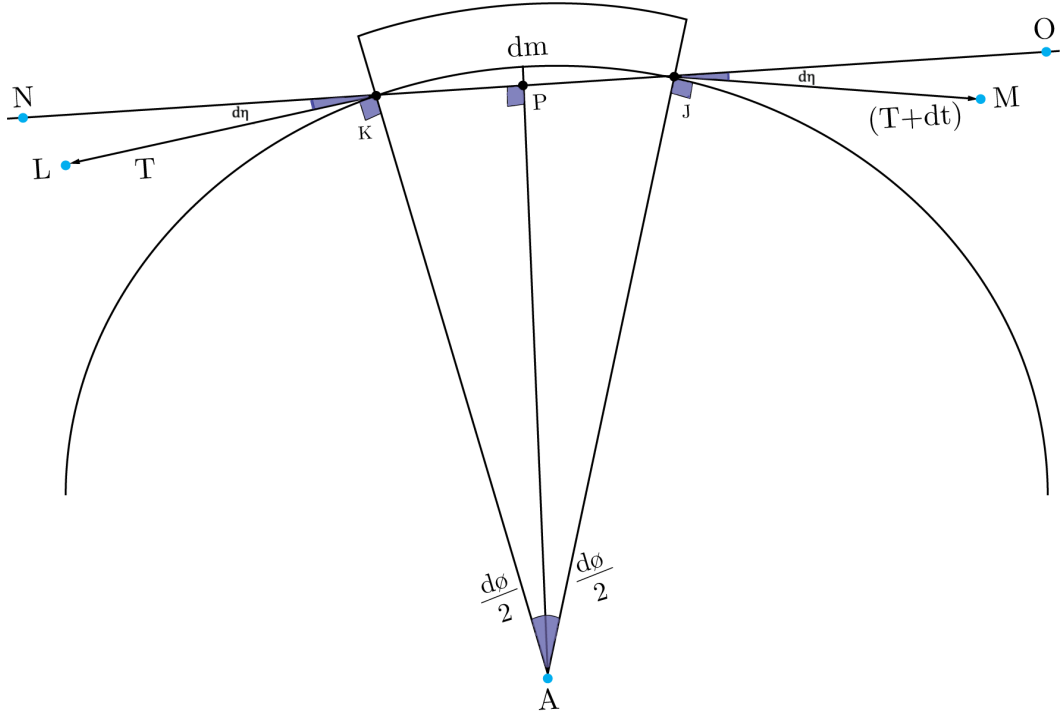

Figure 15: As J is pi/2, along with the point of intersection of NO, $d\eta = \frac{d\phi}{2}$

Substituting $d\eta = \frac{d\phi}{2}$ in equation (11),

$$dm \cdot \frac{d^2y}{dt^2} = -dT + 2\mu_d T \cdot \frac{d\phi}{2}$$

$$\rightarrow dm \cdot \frac{d^2y}{dt^2} = -dT + \mu_d T \cdot d\phi$$

...(12)

To further simplify equation 12, $dm$ can be substituted by $\lambda \cdot dl$ (where $\lambda$ is the linear density of the string and $dl$ is the infinitesimal length of $dm$). When input:

$$\lambda \cdot dl \cdot \frac{d^2y}{dt^2} = -dT + \mu_d T \cdot d\phi$$

Since $dl = r_r \cdot d\phi$ (where $r_r$ is the radius of the rod),

$$\lambda r_r \cdot d\phi \cdot \frac{d^2y}{dt^2} = -dT + \mu_d T \cdot d\phi$$

Rearranging gives,

$$d\phi \left(-\lambda r_r \cdot \frac{d^2y}{dt^2} + \mu_d T\right) = dT$$

$$\frac{dT}{d\phi} = -\lambda r_r \cdot \frac{d^2y}{dt^2} + \mu_d T$$

...(13)

In this differential equation, we are looking at a specific moment in time and hence the acceleration ($\frac{d^2y}{dt^2}$) can be treated as a constant. $\frac{dT}{d\phi}$ demonstrates how the tension in the string varies along the length of the string (varying the angle that $dm$ subtends).

### 2.3.2.4 Solving Differential Equations

Rearranging tension T terms onto one side,

$$\frac{dT}{d\phi} - \mu_d T = -\lambda r_r \cdot \frac{d^2y}{dt^2}$$

Using the integrating factor method, wherein we multiply both sides of the equation with a factor which is found by:

$$e^{\int -\mu_d T \, dx}$$

Which is $e^{-\mu_d \phi}$.

Multiplying both sides with the factor.

$$e^{-\mu_d \phi} \cdot \frac{dT}{d\phi} - e^{-\mu_d \phi} \cdot \mu_d T = -\lambda r_r \cdot e^{-\mu_d \phi} \cdot \frac{d^2y}{dt^2}$$

Since $\frac{d(e^{-\mu_d \phi} \cdot T)}{d\phi} = e^{-\mu_d \phi} \cdot \frac{dT}{d\phi} - e^{-\mu_d \phi} \cdot \mu_d T$ by the product rule,

$$\frac{d(e^{-\mu_d \phi} \cdot T)}{d\phi} = -\lambda r_r \cdot e^{-\mu_d \phi} \cdot \frac{d^2y}{dt^2}$$

Rearrangement yields,

$$d(e^{-\mu_d \phi} \cdot T) = \left(-\lambda r_r \cdot e^{-\mu_d \phi} \cdot \frac{d^2y}{dt^2}\right) \cdot d\phi$$

Integrating both sides gives,

$$\int d(e^{-\mu_d \phi} \cdot T) = \int \left(-\lambda r_r \cdot e^{-\mu_d \phi} \cdot \frac{d^2y}{dt^2}\right) \cdot d\phi$$

$$\rightarrow \int d(e^{-\mu_d \phi} \cdot T) = -\lambda r_r \cdot \frac{d^2y}{dt^2} \int e^{-\mu_d \phi} \cdot d\phi$$

$$e^{-\mu_d \phi} \cdot T = \frac{\lambda r_r}{\mu_d} \cdot \frac{d^2y}{dt^2} \cdot e^{-\mu_d \phi} + C$$

Where C is the constant of integration from indefinite integration.

Dividing both sides by $e^{-\mu_d \phi}$,

$$T = \frac{\lambda r_r}{\mu_d} \cdot \frac{d^2y}{dt^2} + \frac{C}{e^{-\mu_d \phi}}$$

$$T = \frac{\lambda r_r}{\mu_d} \cdot \frac{d^2y}{dt^2} + Ce^{\mu_d \phi}$$

...(14)

### 2.3.2.5 Setting up Boundary Conditions

To find the constant of integration in equation 14, the two known values of Tension -$T_r$ and $T_v$ can be used as boundary conditions. ø is defined to be 0 when T=$T_r$ and $\sigma$ when T=$T_v$. (Figure 16)

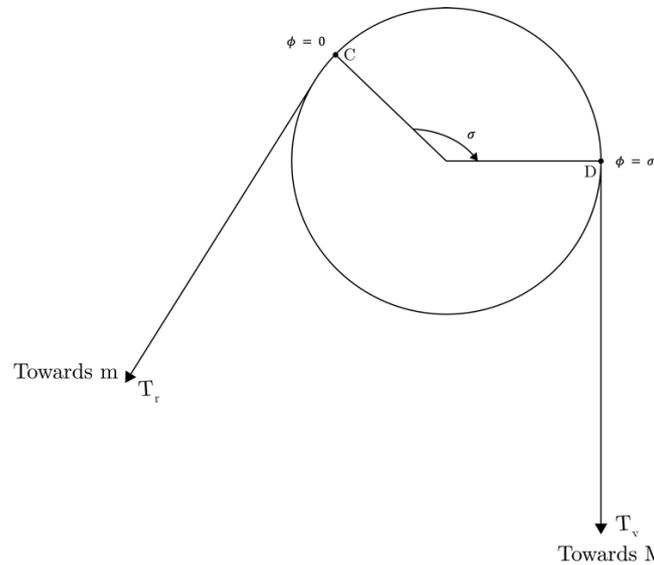

Figure 16: The angle of wrapping $\sigma$ is defined from C to D. At C T = $T_r$ and Ø = 0. While $\sigma$ = ø at T = $T_v$, at D.

Putting ø = 0 and T = $T_r$ in equation (14)

$$T_r = \frac{\lambda r_r}{\mu_d} \cdot \frac{d^2 y}{dt^2} + C e^{\mu_d \cdot 0}$$

$$\Rightarrow T_r = \frac{\lambda r_r}{\mu_d} \cdot \frac{d^2 y}{dt^2} + C \cdot 1$$

Thus,

$$C = T_r - \frac{\lambda r_r}{\mu_d} \cdot \frac{d^2 y}{dt^2}$$

Substituting this expression for C back into the original equation 14.

$$T = \frac{\lambda r_r}{\mu_d} \cdot \frac{d^2 y}{dt^2} + \left(T_r - \frac{\lambda r_r}{\mu_d} \cdot \frac{d^2 y}{dt^2}\right) e^{\mu_d \phi}$$

Rearranging and simplifying:
$$T = -\frac{\lambda r_r}{\mu_d} \cdot \frac{d^2 y}{dt^2} (e^{\mu_d \phi} - 1) + T_r \cdot e^{\mu_d \phi}$$

Since the tension in the string is equal to $T_v$ when ø is equal to $\sigma$:

$$T_v = -\frac{\lambda r_r}{\mu_d} \cdot \frac{d^2 y}{dt^2} (e^{\mu_d \sigma} - 1) + T_r \cdot e^{\mu_d \sigma}$$

...(15)

### 2.3.3 Deriving M's Acceleration $\frac{d^2 y}{dt^2}$

#### 2.3.3.1 Establishment of an Equation for $\frac{d^2 y}{dt^2}$

Substituting expression 15 into 8:

$$\frac{d^2 y}{dt^2} = \frac{T_v}{M} - g$$

$$\frac{d^2 y}{dt^2} = \frac{1}{M}\left(-\frac{\lambda r_r}{\mu_d} \cdot \frac{d^2 y}{dt^2}(e^{\mu_a \sigma} - 1) + T_r \cdot e^{\mu_a \sigma}\right) - g$$

$$\frac{d^2y}{dt^2} = -\frac{\lambda r_r}{M\mu_d} \cdot \frac{d^2y}{dt^2}(e^{\mu_a \sigma} - 1) + \frac{T_r}{M} \cdot e^{\mu_a \sigma} - g$$

...(16)

Since equation (16) requires an expression for $T_r$ like equation (7), the radial motion of $m$ had to be reinvestigated.

### 2.3.3.2 Obtaining an expression for $T_r$

Rewriting equation 7,

$$ma_r = mg \cdot \cos(\theta) - T_r$$

Here, the radial acceleration ($a_r$) is composed of the translational and centripetal acceleration ($a_t$ and $a_c$ respectively).

Since $m$ and $M$ are connected by a rigid string which cannot stretch, both must move with the same translational acceleration. Therefore, $m$'s translational acceleration $a_t$ must be equal to M's ($\frac{d^2y}{dt^2}$).

Since the centripetal of a rotating body is $r\omega^2$ (where r here is defined as l), $a_c = -l\omega^2$ (the negative sign has been defined as radially inwards earlier). Thus[2]

$$a_r = \frac{d^2y}{dt^2} - l\omega^2$$

...(17)

Substituting (17) into (7),

$$m\left(\frac{d^2y}{dt^2} - l\omega^2\right) = mg \cdot \cos(\theta) - T_r$$

Rearranging gives,

$$T_r = -m\frac{d^2y}{dt^2} + mg \cdot \cos(\theta) + ml\omega^2$$

...(18)

---

[2] Here, the signs of the two components of $a_r$ might seem confusing since they are of opposite signs. However, for the most part, since $M$ will be accelerating downward, $\frac{d^2y}{dt^2}$ would take the negative value, and therefore both the components would carry a negative sign.

### 2.3.3.3 Substitution of expression for Tension

Substituting (18) into (16),

$$\frac{d^2y}{dt^2} = -\frac{\lambda r_r}{M\mu_d} \cdot \frac{d^2y}{dt^2}(e^{\mu_d\sigma} - 1) + \frac{\left(-m \cdot \frac{d^2y}{dt^2} + mg \cdot cos(\theta) + ml\omega^2\right)}{M} \cdot e^{\mu_d\sigma} - g$$

Simplifying,

$$\frac{d^2y}{dt^2} = -\frac{\lambda r_r}{M\mu_d} \cdot \frac{d^2y}{dt^2}(e^{\mu_d\sigma} - 1) - \frac{m}{M} \cdot e^{\mu_d\sigma} \cdot \frac{d^2y}{dt^2} + \frac{mg \cdot cos(\theta) + ml\omega^2}{M} \cdot e^{\mu_d\sigma} - g$$

Moving $\frac{d^2y}{dt^2}$ terms onto one side,

$$\frac{d^2y}{dt^2}\left(1 + \frac{\lambda r_r}{M\mu_d} \cdot e^{\mu_d\sigma} - \frac{\lambda r_r}{M\mu_d} + \frac{m}{M} \cdot e^{\mu_d\sigma}\right) = \frac{mg \cdot cos(\theta) + ml\omega^2}{M} \cdot e^{\mu_d\sigma} - g$$

$$\rightarrow \frac{d^2y}{dt^2}\left(\frac{M\mu_d + \lambda r_r \cdot e^{\mu_d\sigma} - \lambda r_r + m\mu_d \cdot e^{\mu_d\sigma}}{M\mu_d}\right) = \frac{mg \cdot cos(\theta) \cdot e^{\mu_d\sigma} + ml\omega^2 \cdot e^{\mu_d\sigma} - Mg}{M}$$

Rearranging and simplifying gives,

$$\frac{d^2y}{dt^2} = \frac{\mu_d mg \cdot cos(\theta) \cdot e^{\mu_d\sigma} + \mu_d ml\omega^2 \cdot e^{\mu_d\sigma} - \mu_d Mg}{M\mu_d + \lambda r_r \cdot e^{\mu_d\sigma} - \lambda r_r + m\mu_d \cdot e^{\mu_d\sigma}}$$

...(19)

Since both $\theta$ and $\sigma$ describe angles, they can be made consistent with one another to simplify equation (19). To do this, we need to derive a relationship between the two.

### 2.3.3.4 Establishing relationship between $\theta$ and $\sigma$

Figure 17 illustrates $\theta$ and $\sigma$ on the same diagram, so a clear distinction between the two can be established.

Figure 17: Since $T_r$ makes an angle $\theta$ with the vertical, its opposite angle ∠CDK = $\theta$

Since Tv and Tr act along tangent to the cylinder at C and D, $\angle \text{ACK} = \angle \text{ADK} = \frac{\pi}{2}$ Thus,

$$\angle \text{CKD} + \angle \text{CAD} = \pi$$

$$\rightarrow \theta + (2\pi - \sigma) = \pi$$

Hence,

$$\sigma = \pi + \theta$$

2.3.3.5 Obtaining final expression for M's acceleration ...(20)

Substituting (20) into (19),

$$\frac{d^2y}{dt^2} = \frac{\mu_d mg \cdot cos(\theta) \cdot e^{\mu_d(\pi+\theta)} + \mu_d ml\omega^2 \cdot e^{\mu_d(\pi+\theta)} - \mu_d Mg}{M\mu_d + \lambda r_r \cdot e^{\mu_d(\pi+\theta)} - \lambda r_r + m\mu_d \cdot e^{\mu_d(\pi+\theta)}} \qquad ...(21)$$

## 2.4 Deriving Expression for l

The length l can be derived by subtracting the length of the string wrapped around the rod ($=r_r\sigma$) and the length of the string moved to the other side (=-y) from the total length of the string (Defined L),

$$l = L + y - r_r\sigma$$

$$\Rightarrow l = L + y - r_r(\pi + \theta) \qquad ...(22)$$

The sign of Y might seem confusing here, but throughout M's motion Y will take a negative value, there L +y would always give a value less than L.

## 2.5 Theory Conclusion

Thus, the following equations describing the motion of the looping pendulum are (6), (21), and (22) respectively.

$$\frac{d^2\theta}{dt^2} = -\frac{3}{2l}g \cdot sin(\theta) \cdot \frac{M_s + 2m}{M_s + 3m} - \frac{2}{l} \cdot \frac{d\theta}{dt}\frac{dl}{dt}$$

$$\frac{d^2y}{dt^2} = \frac{\mu_d mg \cdot cos(\theta) \cdot e^{\mu_d(\pi+\theta)} + \mu_d ml\omega^2 \cdot e^{\mu_d(\pi+\theta)} - \mu_d Mg}{M\mu_d + \lambda r_r \cdot e^{\mu_d(\pi+\theta)} - \lambda r_r + m\mu_d \cdot e^{\mu_d(\pi+\theta)}}$$

$$l = L + y - r_r(\pi + \theta)$$

## 2.6 Numerical Solution using MATLAB

We use MATLAB to numerically integrate the equations and solve them.

### 2.6.1 Code Snippet

Below given is the first portion of the MATLAB code that has been scripted to solve the differential equations. The code is explained entirely in the following section.

```
function dYdt = odefun(t, Y, Ms, m, Mu_d, g, lambda, rr, L, omega,M)
    theta = Y(1);
    dtheta = Y(2);
    y = Y(3);
    dy = Y(4);
    %over here, the second order differential equations have been
    %represented as a vector where the first index is the angular position,
    %second as its angualar acceleration (with respect to time), and the
    %same for the y position (vertical position and acceleration)

    %below all the functions have been listed out, with which the ode45 in
    %built matlab function solving these equations

    % Calculate l
    l = L + y - rr*(3.14159 + theta);

    % Calculate dl_dt
    dl_dt = dy - rr*dtheta;

    % First equation
    d2theta = (-3/(2*l)) * g * sin(theta) * ((Ms + 2*m)/(Ms + 3*m)) - (2/l) *
dtheta * dl_dt;

    % Second equation
    d2y = (Mu_d*m*g*cos(theta)*exp(Mu_d*(pi+theta)) +
Mu_d*m*l*omega^2*exp(Mu_d*(pi+theta)) - Mu_d*M*g) / ...
          (M*Mu_d + lambda*rr*exp(Mu_d*(pi+theta)) - lambda*rr +
m*Mu_d*exp(Mu_d*(pi+theta)));

    dYdt = [dtheta; d2theta; dy; d2y];
    %avighna daruka, ST YAU 2024.
end
```
Verbatim 1: Defining the ODEFUN function in MATLAB.

In MATLAB, the following syntax is used as convention to define a function.

```
function [outputArg1,outputArg2] = untitled(inputArg1,inputArg2)
outputArg1 = inputArg1;
outputArg2 = inputArg2;
end
```
Verbatim 2: Defining a general function in MATLAB.

For the simulation, using this convention, I defined a function called `odefun` with the input arguments being the variables that are being used in the differential equations. dYdt represents the derivatives of the state variables at a given time t. These derivatives are then used by the ode45 solver (a built in MATLAB ode solver) to integrate the system of ODEs over time. The state variables are extracted from the 'Y' vector, which stores the current values of the state variables: [theta, dtheta, y, dy]. After which, I wrote the differential equations that will be solved using the MATLAB ode45 solver. At the end of the script, the revised value of the dYdt function is written after the equations have been solved.

```matlab
% Defining the parameters
Ms = 0.0004; % Mass of string
m = 0.0039; % Mass of moving part (lighter bob m)
Mu_d = 0.257; % Friction coefficient (dynamic friction)
g = 9.80665; % Gravity (constant)
lambda = 0.00059; % Linear density of the string
rr = 0.0003; % Radius of the cylindrical rod
L = 0.500; % Length of the moving bob from the rod
M = 0.030732; % Mass of heavier pendulum bob
Mm_ratio = M / m; % Mass ratio of heavier and lighter mass
omega=1.5;
% Setting initial conditions
theta0 = 1.57079632679; % Initial angle
dtheta0 = 0; % Initial angular velocity
y0 = 0; % Initial y position
dy0 = 0; % Initial y velocity
X0 = [theta0; dtheta0; y0; dy0]; % Initial conditions vector

% Set time span
tspan = [0 0.37]; % Simulation time (changing as mentioned in the paper)

% Solve ODE using ode45 (built in matlab function to solve the odes)
options = odeset('RelTol', 1e-8, 'AbsTol', 1e-8);
[t, X] = ode45(@(t,X) odefun(t, X, Ms, m, Mu_d, g, lambda, rr, L, omega, M), tspan, X0, options);

% Extract solutions
theta = X(:,1);
y = X(:,3);

% Calculate the position of the lighter mass using trignometric functions -
l = L + y - rr * (pi + theta);
x = l .* sin(theta);
y_mass = -l .* cos(theta);

% Calculate the vertical displacement of the heavier bob (for the research
% question)
y_heavy = y;

% Save coordinates to file (so that it can be imported into excel)
coordinates = [x, y_mass];
writematrix(coordinates, 'coord.xlsx');
y_final = y(end); % Final vertical position of the lighter mass (final valye
that was used in the data table)
```

```
heavier_mass_coordinates = [t, y_heavy];
writematrix(heavier_mass_coordinates, 'heaviermass.xlsx');
% Plot the trajectory of the lighter mass
figure;
subplot(2, 1, 1); % Create a subplot for the lighter mass
plot(x, y_mass);
title('Trajectory of Lighter Mass');
xlabel('x (m)');
ylabel('y (m)');
axis equal;
hold on;
plot(x(1), y_mass(1), 'ro', 'MarkerSize', 10);
plot(x(end), y_mass(end), 'go', 'MarkerSize', 10);
plot([0, x(end)], [0, y_mass(end)], 'k--');
legend('Trajectory', 'Start', 'End', 'Final Rod Position', 'Location', 'best');
xlim([-1 1]);
ylim([-0.5 0.3]);

% Plot the vertical displacement of the heavier mass
subplot(2, 1, 2); % Create a subplot for the heavier mass
plot(t, y_heavy);
title('Vertical Displacement of Heavier Mass');
xlabel('Time (s)');
ylabel('Vertical Displacement (m)');
grid on;

% Display final values
fprintf('Final theta: %.4f rad\n', theta(end));
fprintf('Final y: %.4f m\n', y(end));
fprintf('Final x position of mass: %.4f m\n', x(end));
fprintf('Final y position of mass: %.4f m\n', y_mass(end));
fprintf('Vertical distance traveled by heavier mass M: %.4f m\n', y_final);

% Print out coordinates
fprintf('\nCoordinates (x, y):\n');
for i = 1:length(x)
    fprintf('(%.4f, %.4f)\n', x(i), y_mass(i));

end

%AVIGHNA DARUKA ST YAU 2024 RESEARCH COMPETITION
```

Verbatim 3: Main Script to solve and plot the results

### 2.6.2 Code Explanation

#### 2.6.2.1 ODEFUN Function

We call a function called `odefun` with the input parameters being passed.[3] The Y variable over here signals a vector wherein, in the main code, we define the initial conditions of the apparatus. `[theta0, dtheta0, y0, and dy0]`. Next, the value for these initial conditions is fetched from the inputted data. I then wrote all the derived differential equations and then returned the value back into the function, this is repeated until the simulation stops.[4]

---

[3] The code over here was generated with the help of *ChatGPT*. With the prompt being "Generate a MATLAB simulation for the following differential equations"

[4] Further explanation on this methodology is done in section 6.2.

**2.6.2.2 Main Code**

We first begin by defining all the parameters that are being used in the experiment.[5] These are collected either from Literature[6], or measured using suitable apparatus. The only parameter that changes are the masses of both bobs. As stated before, we define the initial conditions of the differential equations and store them in a vector named Y.

The equations are then solved using the ode45 numerical integration solver in MATLAB and the solutions are extracted and stored in an excel file so they can be plotted.

The 30$^{th}$ line in the main script calculates the effective length of the pendulum using the derived equation. The next line calculates the horizontal position of the lighter mass using the effective length that was just calculated (that's why the line is after line 30). Line 32 calculates the vertical position of the lighter mass. Everything is then plotted, but the main coordinates are stored in an excel file, which were copied and then plotted separately on Logger pro. (19)

## 3.0 Preliminary experimentation

As shown in figure 18, initial experimentation was done to help deliver a laboratory plausible research question with which the theory was tested with.

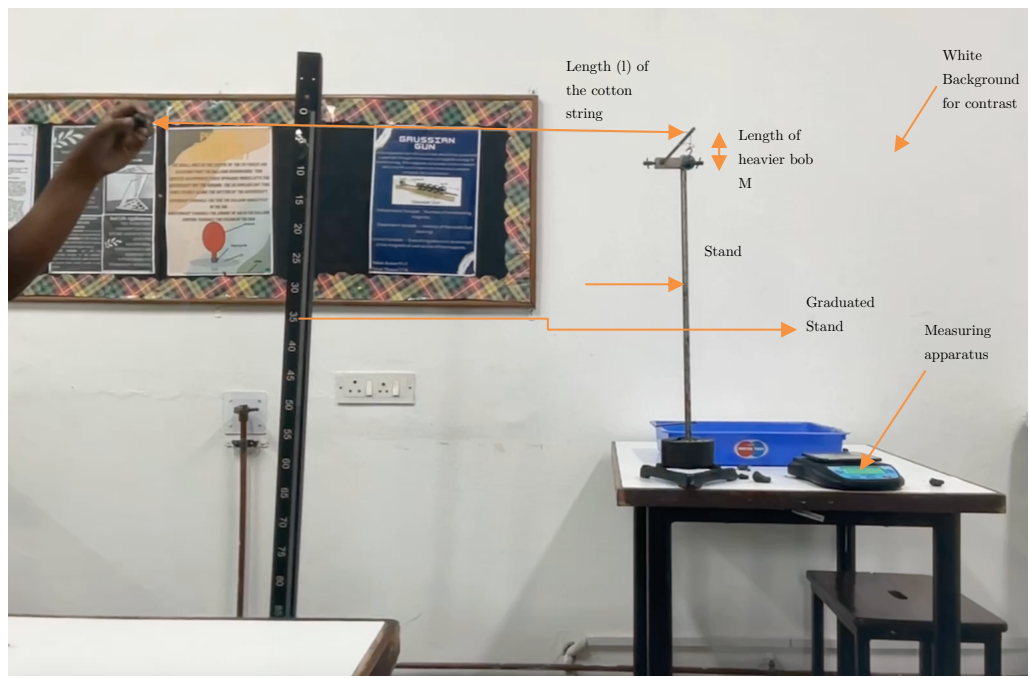

Figure 18: The apparatus used for the preliminary investigation to help determine suitable variables for the experiment.

---

[5] The code over here is used for the mass ratio of 7.88
[6] Only the coefficient of dynamic friction is collected from literature

## 3.1 Possible Independent and Dependent Variables

After a brief investigation, the following independent and dependent variables were chosen as possible candidates.

| Independent Variables | Dependent Variables |
|---|---|
| Mass Ratio of heavier and lighter bob | No. of rotations of the lighter mass as sweeping |
| Initial Angle of release of lighter bob $m$ | Time to stop for heavier bob |
| Initial length of the lighter bob suspended | Time to stop for lighter bob |
| Linear Density of the cotton string | Distance travelled by heavier bob (at vertical) |
| Mass of string used ($M_s$) | Final length of the string of either $m$ or M |
| Coefficient of dynamic friction of the string | Angular velocity of lighter mass |
|  | Average speed of heavier bob $M$. |

## 3.2 Determining independent variable

With the apparatus at hand, the easiest parameter to measure would be the *mass ratio of the heavier and lighter bob,* furthermore, other parameters such as initial length, cause unwanted errors in the experiment. A longer initial length of the string seems to exacerbate the initial bend in the string as shown in figure 19. This leads to experimental inaccuracy.

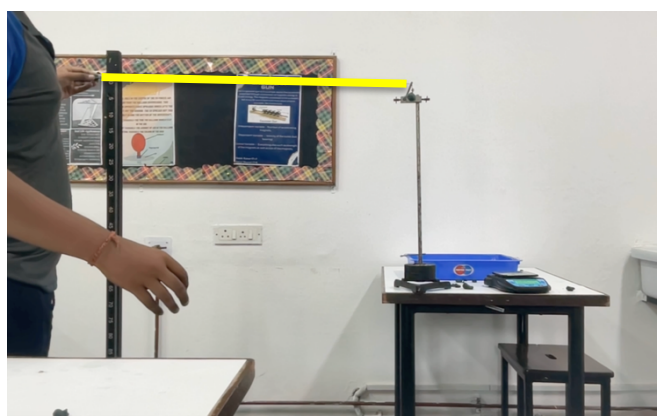 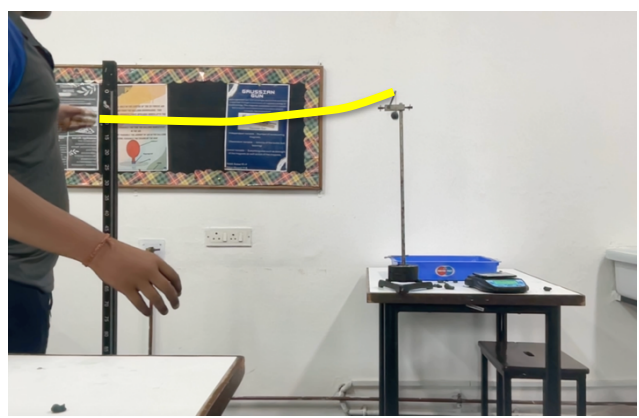

Figure 19(a): Initially, the string is parallel to the ground, as shown by the highlighted yellow line tracing its path.

Figure 19(b): The moment the string is released, there is an initial bend which is visible. This is more pronounced with longer initial lengths.

And hence, the mass ratio of both $M$ and $m$ was chosen as the independent variable.

## 3.3 Determining the dependent variable

The dependent variable was chosen based on the property with the least uncertainty. After the preliminary trials, the distance travelled by the heavier bob was the most accurate over trials taken. And hence, based on pure choice, this was selected for the complete experimentation.

## 4.0 Final RQ for Complete Experimentation

After the preliminary trials, the chosen research question is:

> "How does the mass ratio of heavier bob $M$, and lighter bob $m$, effect the vertical distance travelled by the heavier pendulum bob $M$?"

## 5.0 Experimental Verification

### 5.1 Experimental Setup and Apparatus

An apparatus was constructed to experimentally measure the mass ratio change with respect to time. Figures 20 and 21 show the experimental equipment and camera view that was used for tracking.

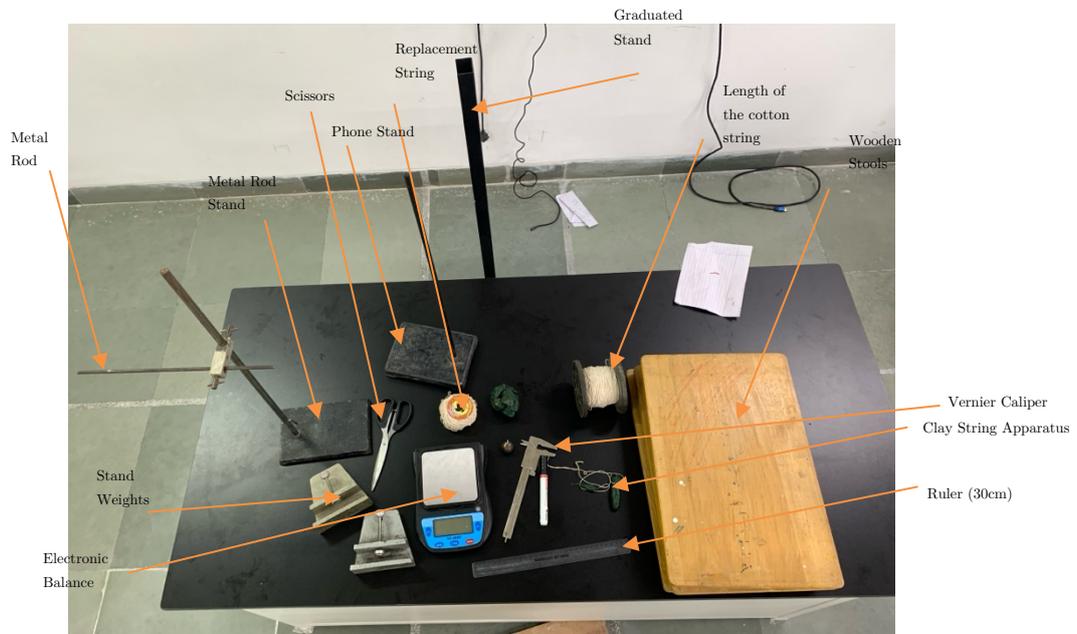
Figure 20: Apparatus Lay out

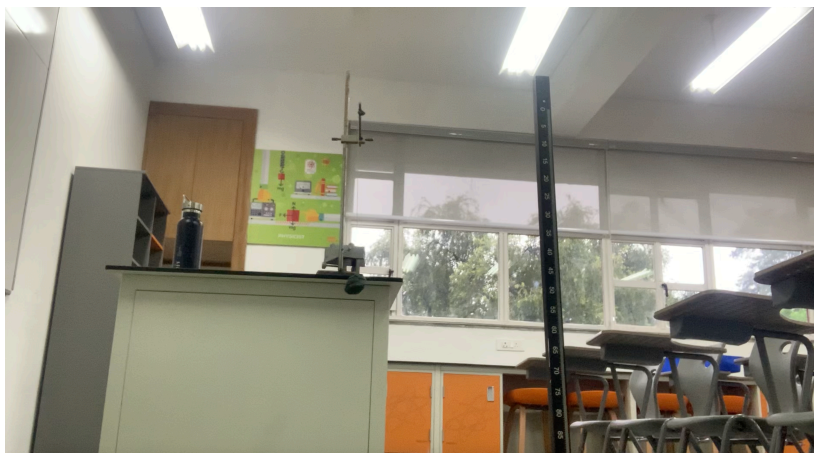
Figure 21: Point of view of the tracking software

During the preliminary trials, it was observed how the string kept moving as it was positioned for release. To counteract this, a very fine indent in the metal rod was made such that the friction between rod and string remains same, but the position of the string would remain constant and not move. This has been shown in Figure 22. Furthermore, to ensure this does not interfere with the accuracy of the experiment, the spot was filed down to make it as frictionless as possible. The rotation of the string around the rod was not interfered because of the divot as its width was exactly the width of the string.

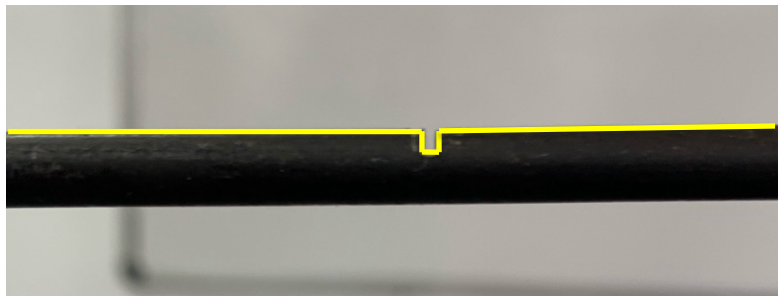

Figure 22: The divot made in the rod to prevent the string from moving.

## 5.2 Data Collection Methodology

To record data, 200 grams of clay was used to model 5 different bob pairs, each with an independent mass ratio. Each pair was connected over a cotton string of linear density 0.00059 $kg \cdot m^{-1}$ with a fixed total length of 0.67 meters, with the calibration distance from bob bobs (used for tracking) being 0.50 meters. The string had a dynamic friction coefficient of 0.257 (11). The entire system rotated around an aluminium rod of radius 0.003 meters, which as mentioned before, has been established as the center of the coordinate system as all tangent points can be averaged out as the center of the rod. This has been visualized in figure 23.

To ensure that the angle of release was consistent throughout trials, a graduated stand with a precision of 0.01m was setup 0.85 meters from the apparatus. At the horizontal, the lighter pendulum bob aligns with the 5cm mark on the graduated stand. This was repeated for every trial, ensuring that the angle is consistent at $\frac{\pi}{2}$ radians.

As mentioned before, the calibration distance has been kept constant at $0.50m$ across trials. 4 different mass ratios gave the range of the independent variable, and six trials were taken for each of these readings.

In respect to video recording, a 30-fps recording device was setup which recorded all the videos from the same frame of reference. This allowed data processing to be more accurate, hence reducing unforced error. Figures 21 and 23 have been taken in this frame of reference. The videos thus obtained were tracked using (Open-Source Physics) Tracker Software.[7] The frame at which the heavier bob seemed to stop moving was observed - to measure the length of the vertical at this frame.

---

[7] One might question that the camera is not positioned perpendicular to the frame of reference, but as this reference frame was consistent throughout all trials – no experimental inaccuracy stemmed from this.

Using the above-described methodology, data collection ensued.

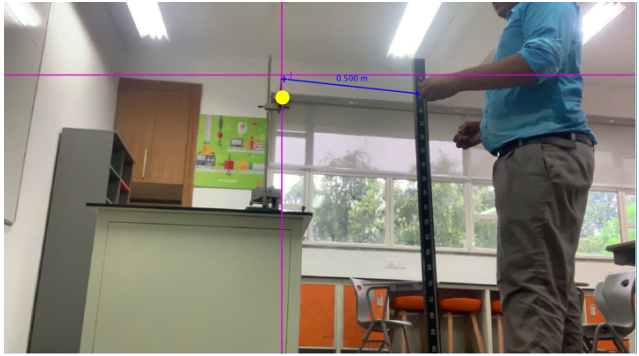

Figure 23(a): The initial position. The length at the vertical aligns with the 5cm mark on the graduated stand. The length has been calibrated at 0.5m as mentioned before.

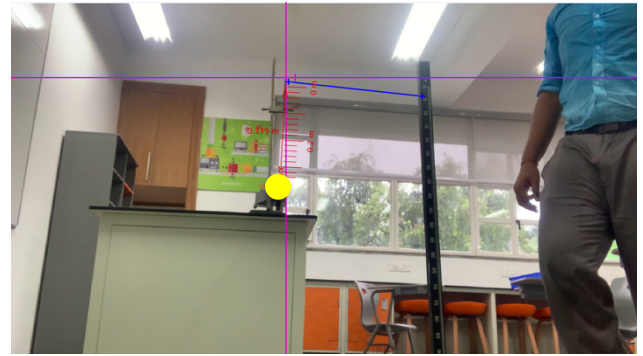

Figure 23(b): The final position. The bob has stopped moving and the length of the vertical has been tracked and measured using the tape measure tool in tracker. Alternatively, to improve accuracy, the -y position can also be noted down as this is parallel to the coordinate axis. **Yellow dot not to scale**

## 6.0 Data Collection and Plotting

### 6.1 Initial Trajectory Comparison

Before the research question was tested, a trajectory comparison was done to ensure that the code was working. Below given is the comparison between simulation trajectory and experimental trajectory that was tracked using the *point mass* tracking tool in tracker. It was realized immediately that the experimental trajectory needed a smoother curve to make a comparison but, with the apparatus at hand, this was not entirely feasible.

The data collected here was done with the mass ratio of 7.88.

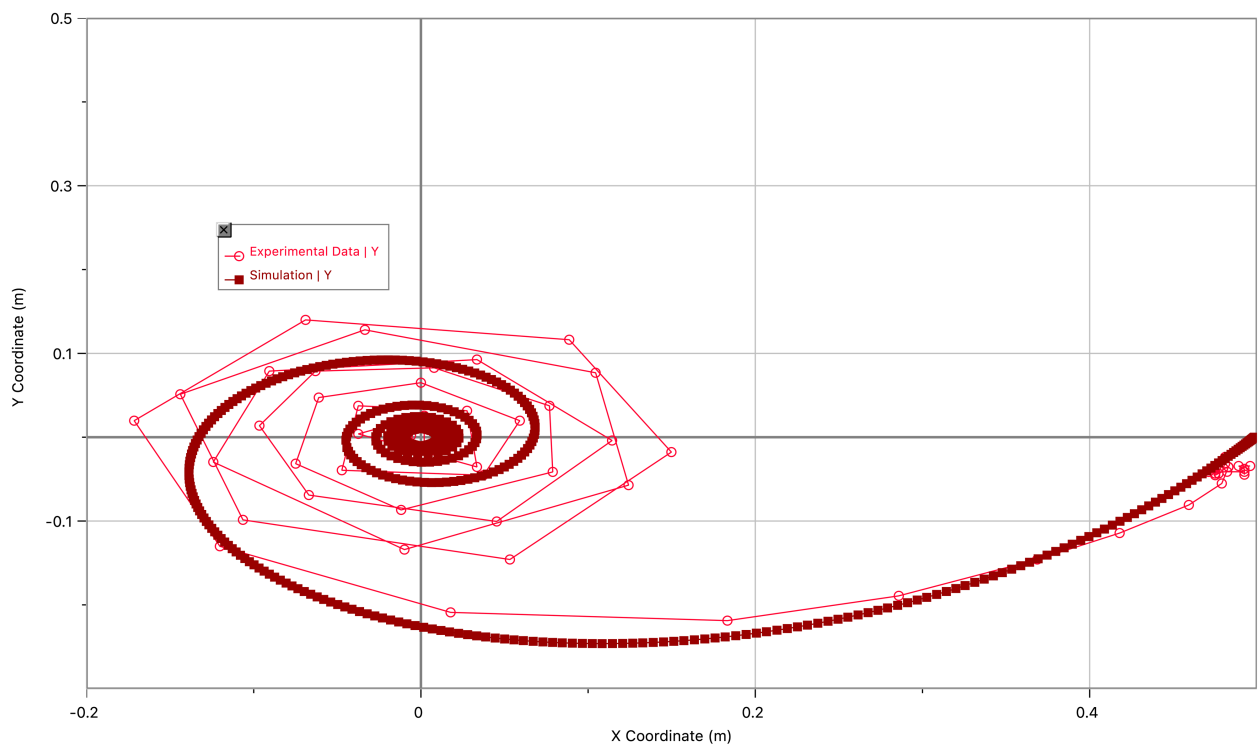

Figure 24: A plot comparison between the theoretical prediction from MATLAB and the experimental data collected using tracker.

## 6.2 Data Collection

Below given is the raw data collected from trials.

| S. No | Mass Ratio(M/m) | Length of the extended string of heavier bob at the vertical (m) | | | | | | Sim. Data (m) | Sim Time Span (s) |
|---|---|---|---|---|---|---|---|---|---|
| | | Trial 1 | Trial 2 | Trial 3 | Trial 4 | Trial 5 | Trial 6 | | |
| 1. | 7.88 | 0.271 | 0.283 | 0.273 | 0.271 | 0.274 | 0.273 | 0.297 | 0.3s |
| 2. | 12.92 | 0.351 | 0.366 | 0.361 | 0.364 | 0.362 | 0.361 | 0.448 | 0.35s |
| 3. | 18.16 | 0.397 | 0.396 | 0.406 | 0.404 | 0.409 | 0.404 | 0.542 | 0.4s |
| 4. | 29.03 | 0.453 | 0.455 | 0.458 | 0.470 | 0.451 | 0.452 | 0.586 | 0.37s |

As shown in the table, for each trial, the simulation run time was different. The start frame of each video and end frame (when the tape measure tool is used to measure length) was determined and the time taken for the pendulum bob to stop moving was used as the time for simulation running in MATLAB. This is a necessity for the MATLAB `odefun` function where we need to define a time span for the solver to solve the differential equation. The data was then processed.

| S. No | Mass Ratio(M/m) | Average length (m) | Uncertainty (MAX-MIN)/2 | Sim. Data (m) |
|---|---|---|---|---|
| 1. | 7.88 | 0.274 | 0.006 | 0.297 |
| 2. | 12.92 | 0.361 | 0.0075 | 0.448 |
| 3. | 18.16 | 0.403 | 0.0065 | 0.542 |
| 4. | 29.03 | 0.457 | 0.0094 | 0.586 |

The data was then plotted.

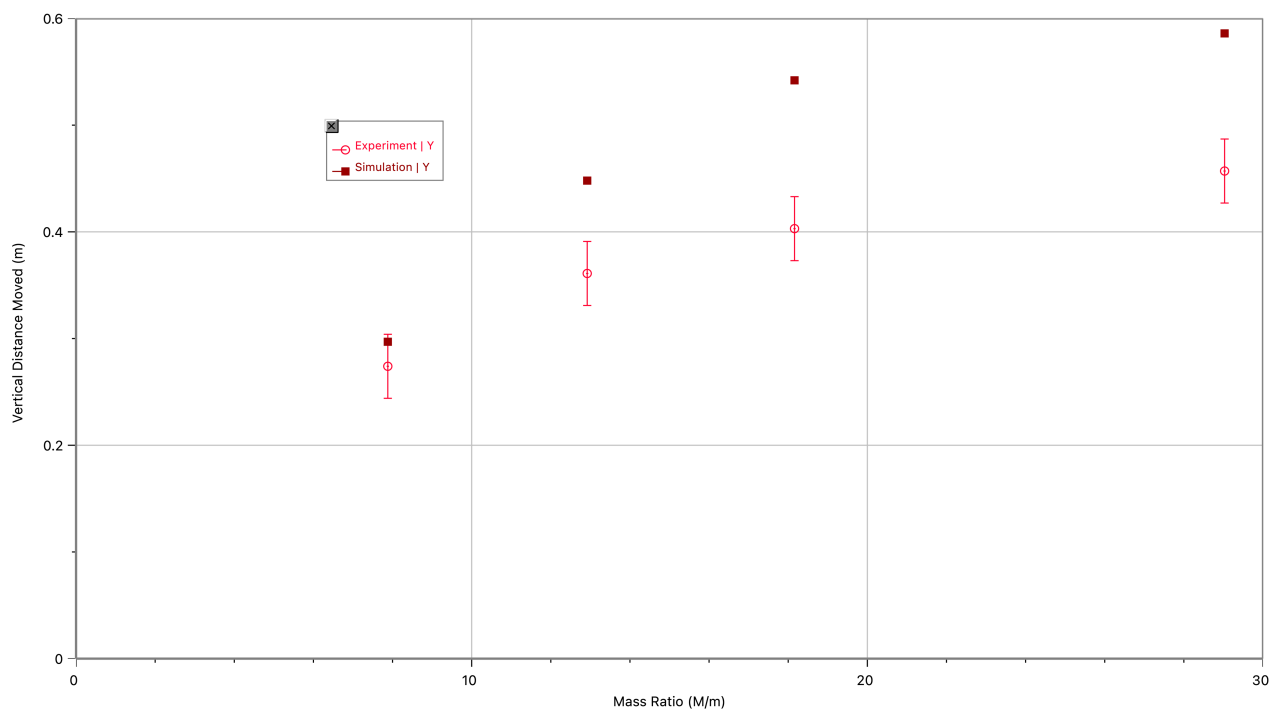

Figure 25: After data processing, the experimental data was compared to the theoretical data. This has been plotted using Logger pro.

### 6.3 Further theoretical verification

Finally, as an extra piece of theoretical verification, the trajectory of the heavier bob $M$ was also compared for both simulation and experiment.

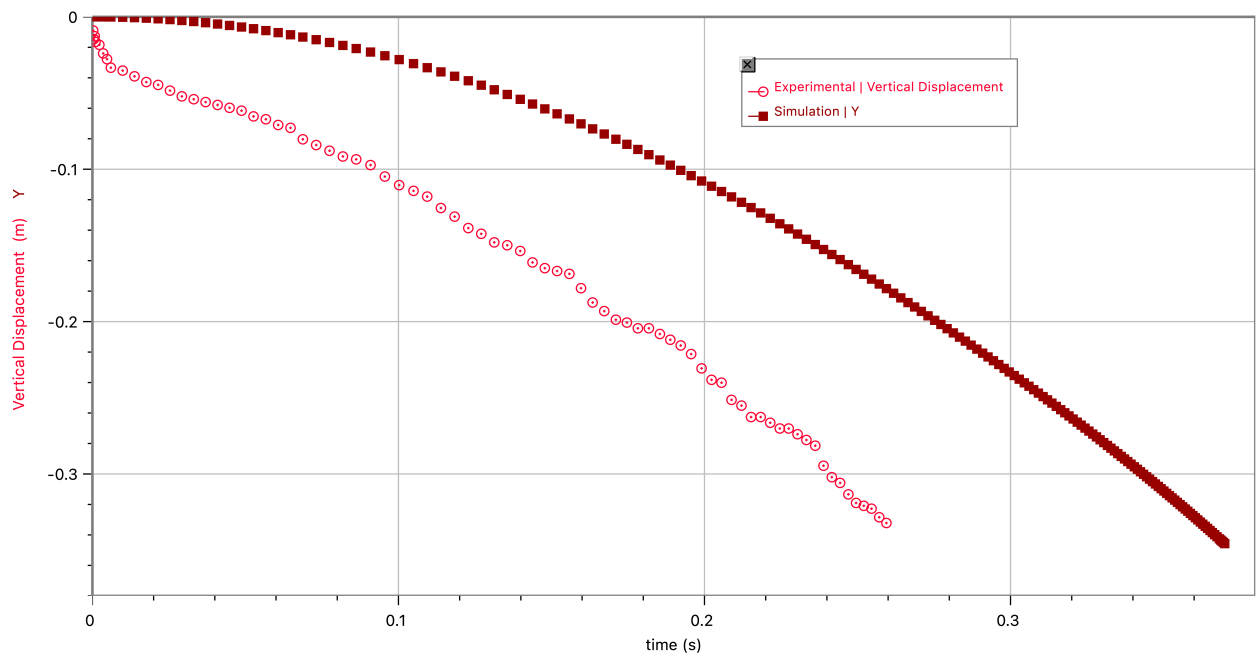

Figure 26: The simulation data and experimental data was also compared for the trajectory of the heavier mass.

As the FPS of the video recording camera tends to a very high number, the experimental data will face a much closer resemblance to the simulation data as the rate at which the displacement *decreases* will obviously be less. But, with the apparatus of a 30 FPS camera, there is not much resemblance with both sets of data.

### 7.0 Data Discussion

From the data collected we can infer that, as the mass ratio grows, so does the vertical distance traveled by the heavier bob in a nonlinear manner. This is understandable when considering the system's dynamics. A higher mass ratio would lead to greater inertia for the heavier bob (as the mass of the lighter bob remains constant). As a result, it is less impacted by tiny forces, allowing for a more stable trajectory, this likely converses into it travelling a greater distance – but for even greater mass ratio this difference decreases, as shown in experimental and simulated data. Furthermore, the lighter mass's influence on the system's motion reduces, causing the heavier bob to experience a more significant vertical displacement.

The modeling and experimental data show similar trends and curve shapes. However, for each beginning angle, the theoretically projected value exceeds the experimental value. The mismatch could be due to string bending, as mentioned in Section 3.2.

Furthermore, the lack of a high FPS camera and limited resources precluded us from taking many readings with different mass ratios. Four trials with six readings each are most likely insufficient to prove a concrete relationship. If this did not match the simulation, the paper's credibility could have been called into question. However, because both the simulation and experimental values follow the same basic shape and trend, it appears that a trend has been established.

## 8.0 Conclusion

From the theoretical derivation, the following differential equations were derived to model the mechanics of the looping pendulum. (6), (21), (22)

$$\frac{d^2\theta}{dt^2} = -\frac{3}{2l} g \cdot sin(\theta) \cdot \frac{M_s + 2m}{M_s + 3m} - \frac{2}{l} \cdot \frac{d\theta}{dt}\frac{dl}{dt}$$

$$\frac{d^2y}{dt^2} = \frac{\mu_d mg \cdot cos(\theta) \cdot e^{\mu_d(\pi+\theta)} + \mu_d ml\omega^2 \cdot e^{\mu_d(\pi+\theta)} - \mu_d Mg}{M\mu_d + \lambda r_r \cdot e^{\mu_d(\pi+\theta)} - \lambda r_r + m\mu_d \cdot e^{\mu_d(\pi+\theta)}}$$

$$l = L + y - r_r(\pi + \theta)$$

After the theory was derived, a simulation was written in MATLAB which solved these equations and plotted the trajectory and vertical displacement of the lighter and heavier bob respectively. This was then compared to the experimental data collected in Tracker.

Figure 25 shows that when the mass ratio increases the vertical length also increases in a nonlinear manner. The experimental values show the same trend as the expected values. However, the former are lower than the latter, which can be attributed to the string's bending.

## 9.0 Evaluation and Error Analysis

### 9.1 Statistical Difference between experimental and simulation results.

At an average, there is a 13% difference between the experimental and simulation results. Although this number is high, it will become lower with lower initial values of the length as the bending of the string as talked about in section 3.2 would reduce with lower values of the length.

With the experimental setup that was employed in the experiment, there should practically be no other factor that effects the values except the bend of the string. To test this, I plotted the difference in experimental and simulation values for each mass ratio.

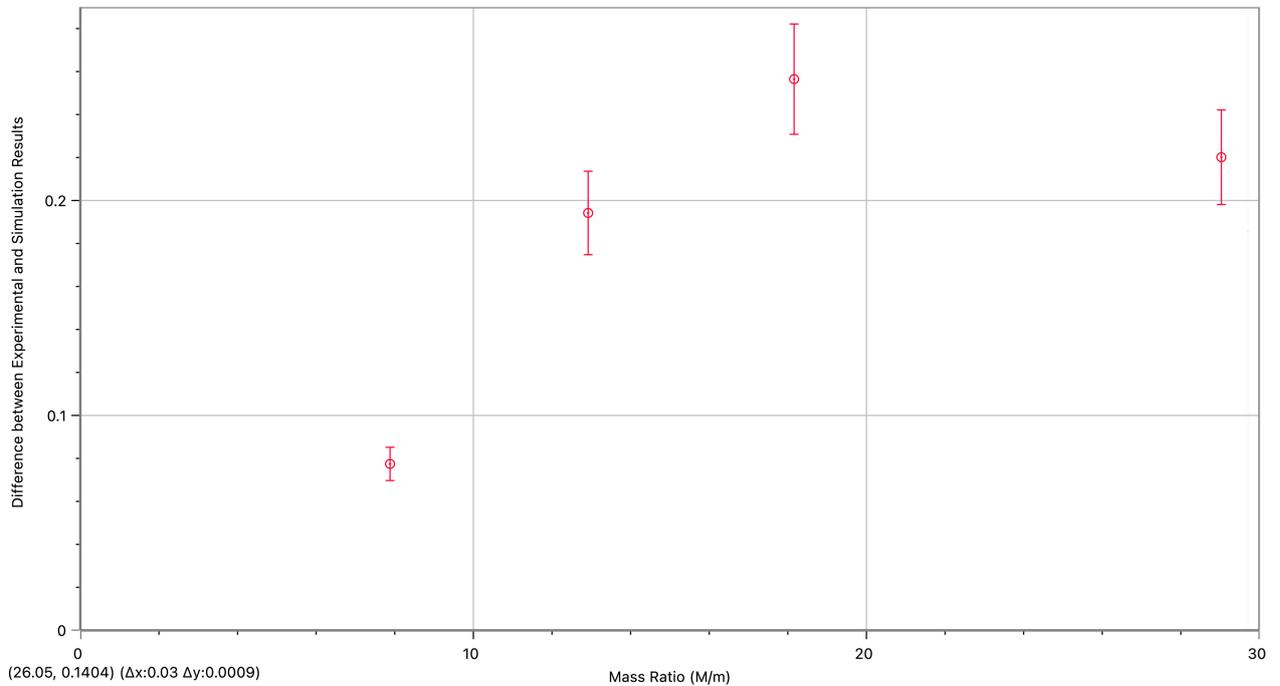

Figure 27: The difference between the experimental and simulation results does not follow any concrete trend.

As shown in the graph, there is no fixed trend for the difference in the experimental and simulation results – this shows how there is random experimental errors and nothing substantial that contributes to the difference between my experiment and the simulation.

Furthermore, as proposed by a paper published before (4) - using a tiny LED instead of clay has improved spiral tracking precision, leading to more precise readings.

My results here do align with papers published previously, but no one has tested this research question in specific. But Reference four's trajectory plot matches exactly with the ones produced here.

## 10.0 Practical Implications

### 10.1 Overview of Practicality

Understanding the mechanics of a looping pendulum has numerous practical applications in our everyday lives and diverse industries. Consider a pendulum in a clock or an amusement park ride that swings back and forth, but with enough energy, it can swing all the way around in a loop. This basic yet effective principle can tell us a lot about how things work.

Engineers, particularly those building amusement park attractions, must understand how a looping pendulum operates. It allows them to develop rides that are both thrilling and safe. Engineers can ensure that rides are sturdy and will not break by understanding the stresses and energy changes involved, making them safer for visitors to enjoy. The code and theory that has been developed here allows anyone to modify the parameters to fit their needs and decode the harmonious mechanics that govern any looping pendulum, whether it be with a cotton string or a massive rotator in an amusement park. Aside from amusement related practical implications, the mechanics of a looping pendulum also play a vital role in timekeeping and navigation. For example, Foucault pendulums – which can be investigated more, indicate the Earth's rotation, use similar concepts. Marine chronometers, which sailors use to determine their longitude at sea, are likewise based on pendulum mechanics.

# 11.0 Appendix

```
function dYdt = odefun(t, Y, Ms, m, Mu_d, g, lambda, rr, L, omega,M)
    theta = Y(1);
    dtheta = Y(2);
    y = Y(3);
    dy = Y(4);
    %over here, the second order differential equations have been
    %represented as a vector where the first index is the angular position,
    %second as its angualar acceleration (with respect to time), and the
    %same for the y position (vertical position and acceleration)

    %below all the functions have been listed out, with which the ode45 in
    %built matlab function solving these equations

    % Calculate l
    l = L + y - rr*(3.14159 + theta);

    % Calculate dl_dt
    dl_dt = dy - rr*dtheta;

    % First equation
    d2theta = (-3/(2*l)) * g * sin(theta) * ((Ms + 2*m)/(Ms + 3*m)) - (2/l) * dtheta * dl_dt;

    % Second equation
    d2y = (Mu_d*m*g*cos(theta)*exp(Mu_d*(pi+theta)) + Mu_d*m*l*omega^2*exp(Mu_d*(pi+theta)) - Mu_d*M*g) / ...
          (M*Mu_d + lambda*rr*exp(Mu_d*(pi+theta)) - lambda*rr + m*Mu_d*exp(Mu_d*(pi+theta)));

    dYdt = [dtheta; d2theta; dy; d2y];
    %avighna daruka, ST YAU 2024.
end
```

```matlab
% Defining the parameters
Ms = 0.0004; % Mass of string
m = 0.0039; % Mass of moving part (lighter bob m)
Mu_d = 0.257; % Friction coefficient (dynamic friction)
g = 9.80665; % Gravity (constant)
lambda = 0.00059; % Linear density of the string
rr = 0.0003; % Radius of the cylindrical rod
L = 0.500; % Length of the moving bob from the rod
M = 0.0212; % Mass of heavier pendulum bob
Mm_ratio = M / m; % Mass ratio of heavier and lighter mass
omega=1.5;
% Setting initial conditions
theta0 = 1.57079632679; % Initial angle
dtheta0 = 0; % Initial angular velocity
y0 = 0; % Initial y position
dy0 = 0; % Initial y velocity
X0 = [theta0; dtheta0; y0; dy0]; % Initial conditions vector

% Set time span
tspan = [0 0.37]; % Simulation time (changing as mentioned in the paper)

% Solve ODE using ode45 (built in matlab function to solve the odes)
options = odeset('RelTol', 1e-8, 'AbsTol', 1e-8);
[t, X] = ode45(@(t,X) odefun(t, X, Ms, m, Mu_d, g, lambda, rr, L, omega, M), tspan, X0, options);

% Extract solutions
theta = X(:,1);
y = X(:,3);

% Calculate the position of the lighter mass using trigonometric functions -
l = L + y - rr * (pi + theta);
x = l .* sin(theta);
y_mass = -l .* cos(theta);

% Calculate the vertical displacement of the heavier bob (for the research
% question)
y_heavy = y;

% Save coordinates to file (so that it can be imported into excel)
coordinates = [x, y_mass];
writematrix(coordinates, 'coord.xlsx');
y_final = y(end); % Final vertical position of the lighter mass (final valye that was used in the data table)

heavier_mass_coordinates = [t, y_heavy];
writematrix(heavier_mass_coordinates, 'heaviermass.xlsx');
% Plot the trajectory of the lighter mass
figure;
subplot(2, 1, 1); % Create a subplot for the lighter mass
plot(x, y_mass);
title('Trajectory of Lighter Mass');
xlabel('x (m)');
ylabel('y (m)');
axis equal;
hold on;
plot(x(1), y_mass(1), 'ro', 'MarkerSize', 10);
plot(x(end), y_mass(end), 'go', 'MarkerSize', 10);
plot([0, x(end)], [0, y_mass(end)], 'k--');
legend('Trajectory', 'Start', 'End', 'Final Rod Position', 'Location', 'best');
xlim([-1 1]);
ylim([-0.5 0.3]);

% Plot the vertical displacement of the heavier mass
subplot(2, 1, 2); % Create a subplot for the heavier mass
plot(t, y_heavy);
title('Vertical Displacement of Heavier Mass');
xlabel('Time (s)');
ylabel('Vertical Displacement (m)');
grid on;

% Display final values
fprintf('Final theta: %.4f rad\n', theta(end));
fprintf('Final y: %.4f m\n', y(end));
fprintf('Final x position of mass: %.4f m\n', x(end));
fprintf('Final y position of mass: %.4f m\n', y_mass(end));
fprintf('Vertical distance traveled by heavier mass M: %.4f m\n', y_final);

% Print out coordinates
fprintf('\nCoordinates (x, y):\n');
for i = 1:length(x)
    fprintf('(%.4f, %.4f)\n', x(i), y_mass(i));

end

%AVIGHNA DARUKA ST YAU 2024 RESEARCH COMPETITION
```

This page has been left blank intentionally